\documentclass[12pt]{article}
\usepackage{latexsym}
\usepackage{amsfonts}

\title{Invariants of the nilpotent and solvable triangular Lie algebras}
\author{S. Tremblay\thanks{Centre de recherches math\'ematiques and
D\'epartement
de physique, Universit\'e de Montr\'eal, C.P. 6128, succ. Centre-ville,
Montr\'eal (QC), H3C 3J7, Canada (\texttt{tremblay@crm.umontreal.ca})} \and
P. Winternitz\thanks{Centre de recherches
math\'ematiques and D\'epartement de math\'ematiques et de statistique,
Universit\'e de Montr\'eal,
C.P. 6128, succ. Centre-ville, Montr\'eal (QC), H3C 3J7, Canada
(\texttt{wintern@crm.umontreal.ca})}}
\date{}
\makeatletter
\@addtoreset{equation}{section}

\makeatletter

\def\neq {\not\equiv}

\def\be   {\begin{equation}}   \def\ee   {\end{equation}}
\def\ba   {\begin{array}}      \def\ea   {\end{array}}
\def\bea  {\begin{eqnarray}}   \def\eea  {\end{eqnarray}}
\def\bean {\begin{eqnarray*}}  \def\eean {\end{eqnarray*}}

\newcommand{\diag}{\mathrm{diag}}

\begin{document}
%\english
\maketitle

\begin{abstract}
Invariants of the coadjoint representation of two classes of Lie algebras are
calculated. The first class consists of the nilpotent Lie algebras $T(M)$,
isomorphic to the algebras of upper triangular $M\times M$ matrices. The Lie
algebra $T(M)$ is shown to have $[M/2]$ functionally independent invariants.
They can all be chosen to be polynomials and they are presented explicitly. The
second class consists of the solvable Lie algebras $L(M,f)$ with $T(M)$ as their
nilradical and $f$ additional linearly nilindependent elements. Some general
results on the invariants of $L(M,f)$ are given and the cases $M=4$ for all $f$
and
$f=1$, or $f=M-1$  for all $M$ are treated in detail.
\end{abstract}

\section{Introduction}
The purpose of this paper is to present some results on the invariants of two
classes
of Lie algebras, over the field of complex or real numbers ($K=\mathbb{C}$ or
$\mathbb{R}$). The first class are the finite triangular nilpotent Lie
algebras
$T(M)$ of dimension $M(M-1)/2$. By triangular nilpotent Lie algebra, we mean the
nilpotent
Lie
algebra isomorphic to the Lie algebra of strictly upper triangular
$M \times M$
matrices. The second class of algebras studied below are the finite solvable
triangular Lie algebras $L(M,f)$ that
have
$T(M)$ as their nilradicals (maximal nilpotent ideals) and contain $f$ further
nonnilpotent
elements. For the algebras $L(M,f)$, use will be made of a recent article
\cite{1} in
which we obtained a classification of such Lie algebras and presented the
general
form of the commutation relations.

In physics, invariant operators of the symmetry group of a physical system and
its
subgroups provide quantum numbers. Indeed, the eigenvalues of the invariant
operators
of the entire symmetry group will be the quantum numbers, characterizing the
system
as such
(e.g., the particle mass and spin in the case of the Poincar\'e group).
The
invariant operators of subgroups will then characterize states of the system
(its
energy, linear or angular momentum, etc.) \cite{2}.

In other applications, invariant operators of  dynamical groups provide mass
formulas
\cite{3,4}, energy spectra \cite{5,6} and in general characterize specific
properties
of physical systems.

Let us stress here that in this context the concept of an invariant need not
mean a
Casimir operator. Indeed, the problem of finding invariants will be reduced to
that
of solving a certain set of linear first order partial differential equations
\cite{7,8}. These may have polynomial solutions, giving rise to Casimir
operators. They may
also
have rational solutions, giving rise to rational invariants. Finally, the
equations may have more general
solutions, including transcendental functions of various types, leading to
general invariants.

Casimir operators are polynomials in the enveloping algebra of a Lie algebra
that
commute with all elements of the Lie algebra. In other words, a Casimir
operator of a Lie algebra is an element of the centre of the enveloping
algebra. For a Lie algebra $L$, the Casimir operators can be calculated
directly. Namely,
we impose that a general polynomial in the enveloping algebra commutes with
all basis elements $X_i$ of $L$. However, more efficiently,
they can be calculated as invariants of the coadjoint representation of the
corresponding Lie algebra \cite{9,10}.

The Casimir operators of semisimple Lie algebras are well known. Their number
$p$ is equal to the rank of the considered Lie algebra \cite[\ldots,17]{11}.
Moreover, for semisimple Lie algebra, all invariants of the coadjoint
representation can be expressed as functions of $p$ homogeneous polynomials.

For solvable Lie algebras, the situation is less clear. Neither the specific
type of functions, nor the number of functionally independent invariants is
known.

One method, for calculating the polynomial and other invariants for arbitrary
Lie
algebras, is an infinitesimal one. This method has been presented in \cite{7}
and
applied to low dimensional Lie algebras \cite{18,19}, to subalgebras of the
Poincar\'e Lie
algebra
\cite{20} and to solvable Lie algebras with Heisenberg or Abelian nilradicals
\cite{21,22}.

From a mathematical point of view, in the representation theory of solvable Lie
algebras, polynomial and non-polynomial invariants in the coadjoint
representation
appear on the same footing: they characterize irreducible representations.
Casimir
operators in the enveloping algebra correspond to polynomial invariants. The
functions of the infinitesimal operators, corresponding to the non-polynomial
invariants, will be called `generalized Casimir operators'. In the study of
the
integrability of classical Hamiltonian systems, integrals of motion do not
have
to be
polynomials in the dynamical variables \cite{23,24}.

In Section 2 we formulate the problem of calculating the invariants of the
coadjoint representation. Section~3 is devoted to the nilpotent algebras $T(M)$.
We calculate the invariants explicitly. There are $[M/2]$ functionally
independent invariants, all of them polynomials. In Section~4 we calculate the
invariants of the solvable Lie algebras $L(M,f)$. We first treat the case $M=4$
in detail, then present results and conjectures for $L(M,M-1)$ and $L(M,1)$.

\section{General results and formulation of the problem}
Let us consider a $N$-dimensional Lie algebra given by the basis
$\{Y_{1},\ldots,Y_{N}\}$ and the commutation relations

\begin{equation}
\left[Y_{i},Y_{j}\right]=\sum_{k=1}^{N}C_{ij}^{k}\,Y_{k}
\,\,\,\,\,\,\,\,\, 1\leq i,j,k\leq N.
\label{eq:YY}
\end{equation}

In order to calculate the invariants of the Lie algebra $L$, we
shall work on the dual of $L$. We consider smooth functions $F:\
(y_{1},\ldots,y_{N})\rightarrow K$ where the variables $y_{i}$ are
ordinary (commuting) variables on the space $L^{*}$, dual of $L$,
and $K$ is the field of complex or real numbers ($K=\mathbb{C}$ or
$\mathbb{R}$). The generators $Y_{i}$ are given in the coadjoint
representation by the differential operators

\begin{equation}
\hat Y_{i}=\sum_{j,k} C_{ij}^{k}\,y_{k}\frac{\partial}{\partial y_{j}}\,.
\label{eq:Yi}
\end{equation}
We can verify easily that the differential operators $\hat Y_{i}$ satisfy the
commutation relations (\ref{eq:YY}).

The function $F$ will be an invariant of the coadjoint representation of $L$
if
it
satisfies the linear first order partial differential equations

\begin{equation}
\hat Y_{i}\cdot F=0 \,\,\,\,\,\,\,\,\, i=1,\ldots, N\;
\label{eq:YF}
\end{equation}
which, one hopes, can be solved by standard methods.

Our aim is to find a complete set of functionally independent solutions to
equation
(\ref{eq:YF}), for nilpotent and solvable triangular Lie algebras. If the
solutions
are polynomials, we obtain Casimir operators by replacing the variables
$y_{i}$
by
the generators $Y_{i}$ and symmetrizing, whenever necessary. The number of
independent solutions $n_I$, i.e. the number of functionally independent
invariants, is equal to

\begin{equation}
n_I=N-rank(M)
\label{eq:ni}
\end{equation}
where $M$ is the antisymmetric matrix with elements

\begin{equation}
M_{ij}=\sum_{k=1}^{N}C_{ij}^{k}\, y_k
\label{eq:Mij}
\end{equation}
(see Ref.\cite{7}).

\section{Invariants of nilpotent triangular Lie algebras}

\subsection{Structure of the nilpotent Lie algebra $T(M)$ and its realization
by differential operators}

Let us consider the finite triangular Lie algebra $T(M)$ over the field $K$ of
complex or real numbers. A basis for this algebra is

\begin{equation}
\begin{array}{c}
\{N_{ik}\ \mid\ 1\leq i<k\leq M\}
\label{eq:base} \\*[2ex]
(N_{ik})_{ab}=\delta_{i,a}\ \delta_{k,b}\ \ \ \ \ \
\dim\ T(M)=\frac{1}{2}M(M-1)\equiv r
\end{array}
\label{eq:base2}
\end{equation}
with $M>3$. The Lie algebra $T(2)$ is trivial and $T(3)$ is
isomorphic to the Heisenberg Lie algebra $H(1)$. The dimension
$M=3$ is the only case for which there is an isomorphism between
the triangular and the Heisenberg Lie algebras.

The commutation relations of $T(M)$ are given by

\begin{equation}
[N_{ik},N_{ab}]=\delta_{k,a} N_{ib}-\delta_{b,i} N_{ak}.
\label{eq:NN}
\end{equation}
This basis can be represented by the standard basis of the strictly upper
triangular
$M\times M$ matrices.

The differential operators $\hat N_{ik}$ realizing the coadjoint
representation
of
$T(M)$, are

\begin{equation}
\hat N_{ik}=\sum_{b=k+1}^{M} n_{ib}\frac{\partial}{\partial
n_{kb}}-\sum_{a=1}^{i-1}
n_{ak}\frac{\partial}{\partial n_{ai}}.
\label{eq:N}
\end{equation}
Note that $\hat N_{1M}\equiv 0$ in (\ref{eq:N}), since $N_{1M}$ commutes with
all
the
elements of $T(M)$.

We shall realize the coadjoint representation of $T(M)$ in a space of
differentiable
functions of $r$ variables, i.e.

\begin{equation}
F=F(n_{12},n_{23},\ldots, n_{(M-1)M}, n_{13},n_{24},\ldots, n_{(M-2)M},\ldots,
n_{1M}).
\label{eq:F(n)}
\end{equation}
The function $F$ will be an invariant of the coadjoint representation of
$T(M)$,
if
it satisfies the linear first order partial differential equations

\begin{equation}
\hat N_{ik}\cdot F=0 \,\,\,\,\,\,\,\,\, 1\leq i<k \leq M.
\label{eq:NF}
\end{equation}

\subsection{Definitions and results}
Let us consider the set of strictly upper triangular $M \times M$ matrices
$Q=Q(M)$
over
the field $K$ i.e.

\begin{equation}
Q_{ik}=\left\{
\begin{array}{lll}
n_{ik} &for\ \ k-i\geq 1 \\
\\
0 &otherwise.
\end{array}
\right.
\label{eq:Q}
\end{equation}
We define the determinant $Z_{\mu}=Z_{\mu}(M)$ constructed from the
$\mu\times \mu$ right
upper corner sub-matrix of the matrix $Q$, i.e.

\begin{equation}
Z_{\mu}=
\left|
\begin{array}{cccc}
n_{1(M-\mu+1)} & n_{1(M-\mu+2)} & \cdots & n_{1M}  \\
n_{2(M-\mu+1)} & n_{2(M-\mu+2)} & \cdots & n_{2M} \\
\vdots & \vdots & & \vdots \\
n_{\mu(M-\mu+1)} & n_{\mu(M-\mu+2)} & \cdots & n_{\mu\,M}
\end{array}
\right|\;\;\;\; 1\leq \mu\leq \left[\frac{M}{2}\right]
\label{eq:Z}
\end{equation}
where we shall use the standard notation $[x]$ for the entire part of a
positive
number. In particular,

\begin{equation}
p=\left[\frac{M}{2}\right]=\left\{
\begin{array}{lll}
\frac{M}{2} &for\ \ M=2p \\
\\
\frac{M-1}{2} &for\ \ M=2p+1.
\end{array}
\right.
\label{eq:p}
\end{equation}

\newtheorem{theo1}{Theorem}
\begin{theo1}
The triangular Lie algebra $T(M)$ defined by equations $(\ref{eq:base})$ and
$(\ref{eq:NN})$ has exactly $[M/2]$ functionally
independent
invariants. A basis of invariants is given by
\begin{equation}
I_{\mu}=Z_{\mu} \,\,\,\,\,\,\,\,\, \mu=1,\ldots,\left[\frac{M}{2}\right]
\end{equation}
where $Z_{\mu}$ is the determinant function given by eq.(\ref{eq:Z}).
\label{theo:I=Z}
\end{theo1}

\emph{Proof.} Let us first consider the cases $M$ odd, i.e. $M=2p+1$ for
$p=2,3,\ldots$

We begin by applying the set of $p(p+2)$ differential operators of
eq.(\ref{eq:N}),
given by
\begin{equation}
\begin{array}{cccc}
\hat N_{1(p+1)} & \hat N_{1(p+2)} & \cdots & \hat N_{1M} \\
\hat N_{2(p+1)} & \hat N_{2(p+2)} & \cdots & \hat N_{2M} \\
\vdots & \vdots & & \vdots \\
\hat N_{p(p+1)} & \hat N_{p(p+2)} & \cdots & \hat N_{pM}\\
0 & \hat N_{(p+1)(p+2)} & \cdots & \hat N_{(p+1)M}
\end{array}
\end{equation}
on the functions (\ref{eq:F(n)}). The action of all these operators eliminates
the
dependence on the $p(p+1)$ variables $n_{ik}$ for

\begin{equation}
1\leq i\leq p\ \ \ \ \ \ i+1\leq k \leq p+1
\end{equation}
and
\begin{equation}
p+1\leq i\leq M-1\ \ \ \ \ \ i+1\leq k \leq M\;.
\end{equation}
The $p^{2}$ remaining variables are

\begin{equation}
\begin{array}{cccc}
n_{1(p+2)} & n_{1(p+3)} & \cdots & n_{1M} \\
n_{2(p+2)} & n_{2(p+3)} & \cdots & n_{2M} \\
\vdots & \vdots & & \vdots \\
n_{p(p+2)} & n_{p(p+3)} & \cdots & n_{pM}
\end{array}
\end{equation}
and the $p(p-1)$ remaining differential operators $\hat N_{ik}$ of
eq.(\ref{eq:N})
are given by

\begin{eqnarray}
&\hat N_{ik}&=\sum_{b=p+2}^{M}n_{ib}\frac{\partial}{\partial n_{kb}}\ \ \ \
1\leq i\leq p-1 \ \ \ \ i+1\leq k \leq p
\label{eq:Na}\\
&\hat N_{ik}&=-\sum_{a=1}^{p}n_{ak}\frac{\partial}{\partial n_{ai}}\ \ \ \
 p+2\leq i\leq M-1 \ \ \ \ i+1\leq k \leq M.
\label{eq:Nb}
 \end{eqnarray}
These differential operators are linearly independent. Therefore, the number
of
invariants for $T(2p+1)$ is $p$, i.e. the difference between the number of
remaining
variables and the number of remaining independent differential operators.

At this stage of the proof it is sufficient to verify that the remaining
differential operators (\ref{eq:Na}) and (\ref{eq:Nb}) annihilate determinants
$Z_{1},\ldots,Z_{p}$, i.e. $\hat N_{ik}\cdot Z_{\alpha}=0$ for
$\alpha=1,\ldots,p$.

Let us first consider the set of differential operators (\ref{eq:Na}). A given
differential operator $\hat N_{ik}$ ($i$ and $k$ fixed) of (\ref{eq:Na})
annihilates
the determinant $Z_{1},\ldots,Z_{k-1}$, since the variables $n_{kb}$ ($p+2\leq
b\leq
M$) do not figure in these determinants. It is therefore sufficient to look
how
$\hat
N_{ik}$ acts on $Z_{k},\ldots,Z_{p}$.

The determinant $Z_{\beta}$ $\beta \in \{k,k+1,\ldots,p\}$ can be expanded in
terms
of its $k^{th}$ row
\begin{equation}
Z_{\beta}=\sum_{b=2p+2-\beta}^{M}n_{kb}\,C_{kb}^{(\beta)}
\label{eq:devZa}
\end{equation}
where $C_{kb}^{(\beta)}$ is the cofactor of the $\beta \times \beta$ square
matrix
associated with the determinant $Z_{\beta}$. Hence, the differential operator
$\hat
N_{ik}$ applied on these determinants gives
\begin{equation}
\hat N_{ik}\cdot Z_{\beta}=\sum_{b=2p+2-\beta}^{M}n_{ib}\,C_{kb}^{(\beta)}.
\label{eq:devNZa}
\end{equation}
The right hand side of eq.(\ref{eq:devNZa}) vanishes, since it corresponds to
the
expansion of determinant in terms of the cofactors of a different row. This
gives the
determinant of a matrix with two identical rows, hence zero.

The procedure is very similar for the set of differential operators
(\ref{eq:Nb}). An operator $\hat N_{ik}$ of this set annihilates the
determinants
$Z_{1},\ldots,Z_{M-i}$ since the operator acts only on variables not figuring in
the determinants. Let us consider the action of $N_{ik}$ in (\ref{eq:Nb}) for
the
determinants $Z_{\gamma}$, where $\gamma \in \{(M-i+1),(M-i+2),\ldots,p\}$.

We can write the determinants $Z_{\gamma}$ as
\begin {equation}
Z_{\gamma}=\sum_{a=1}^{\gamma}n_{ai}\,C_{ai}^{(\gamma)}
\label{eq:deta}
\end{equation}
and the action of the differential operators $\hat N_{ik}$ on these
determinants
is
given by

\begin{equation}
\hat N_{ik}\cdot Z_{\gamma}=-\sum_{a=1}^{\gamma}n_{ak}\,C_{ai}^{(\gamma)}.
\end{equation}
Hence, we obtain a determinant with two identical columns. More precisely, the
action
of
differential operator $\hat N_{ik}$ in (\ref{eq:Nb}) on determinants
(\ref{eq:deta})
is the following: the column $n_{ai}$ in determinants $Z_{\gamma}$ is replaced
by the
column $-n_{ak}$, for $1\leq a \leq \gamma$. Therefore, by the property of
determinants, this action annihilates $Z_{\gamma}$.

The proof for the even case is very similar to the odd case and we omit
it.
$\Box$

\section{Invariants of the solvable triangular Lie algebras}
\subsection{Structure of the solvable triangular Lie algebra $L(M,f)$}

In this section we sum up the main results of Ref.\cite{1} to make this article
self-contained.

Let us extend the algebra $T(M)$ to an indecomposable solvable Lie algebra
$L(M,f)$
of dimension $d=\frac{1}{2}M(M-1)+f$ having $T(M)$ as its nilradical. In
other
words, we add $f$ further linearly nilindependent elements to $T(M)$. Let us
denote
them $\{X^{1},\ldots,X^{f}\}$.

\newtheorem{def1}{Definition}
\begin{def1}
$\bullet$ A set of elements $\{X^{\alpha}\}$ of a Lie algebra $L$ is linearly
nilindependent if no nontrivial linear
combination of them is a nilpotent element.

$\bullet$ A set of matrices $\{A^{\alpha}\}_{\alpha=1,\ldots,n}$ is linearly
nilindependent if no nontrivial linear
combination of them is a nilpotent matrix, i.e. if
\begin{eqnarray}
\left(
\sum_{i=1}^{n} c_{i}\,A^{i}
\right)^{k}=0
\end{eqnarray}
for some $k\in \mathbb{Z^{+}}$, implies $c_{i}=0\;\;\forall i$.
\label{eq:nilp}
\end{def1}

The results on the structure of the Lie algebras $L(M,f)$ that we have
obtained
in
\cite{1} can be summed up as follows.

Each Lie algebra $L(M,f)$ can be transformed to a canonical basis
$\{X^{\alpha},N_{ik}\}\ ,\ \alpha=1,\ldots,f\ ,\ 1\leq i<k\leq M\,$ with
commutation
relations (\ref{eq:NN}) and

\begin{eqnarray}
[X^{\alpha},N_{ik}] &=& \sum_{p<q}A^{\alpha}_{ik\,,\,pq}\ N_{pq}\label{eq:xn}
\\*[2ex]
[X^{\alpha},X^{\beta}] &=& \sigma^{\alpha\beta} N_{1M}\label{eq:xx} \\*[2ex]
\nonumber
1\leq\alpha,\beta\leq f  && A^{\alpha}_{ik\,,\,pq},\ \sigma^{\alpha\beta}\in K.
\end{eqnarray}
The commutation relations (\ref{eq:xn}) can be rewritten as
\begin{equation}
\begin{array}{c}
[X^{\alpha},N]=A^{\alpha}N \\*[2ex]
N\equiv(N_{12}\ N_{23}\ldots N_{(M-1)M}\ N_{13}\ldots N_{(M-2)M}\ldots
N_{1M})^{T}\\*[2ex]
A^{\alpha}\in K^{r\times r}\ \ \ \ N\in K^{r\times 1}
\end{array}
\label{eq:xamatrice}
\end{equation}
where the superscript $T$ indicates transposition. We mention that the
vector
$N$
introduces an order in lines (columns) of the matrices $A^{\alpha}$, where
each
line
(column) is represented by two numbers. The matrices
$A^{\alpha}=\{A^{\alpha}_{ik\,,\,pq}\}$ have the following canonical form.

\begin{description}
\item[(i)] They are upper triangular.
\item[(ii)] The only off-diagonal matrix elements that do not vanish
identically
and
cannot be annulled by a redefinition of the elements $X^{\alpha}$ are:
\begin{equation}
A_{12\,,\,2M}^{\alpha}\ \ \ \ A_{j(j+1)\,,\,1M}^{\alpha}\ (2\leq j\leq M-2)\
 \ \ \ A_{(M-1)M\,,\, 1(M-1)}^{\alpha}.
\label{eq:nonzero}
\end{equation}
\item[(iii)] The diagonal elements $a^{\alpha}_{i(i+1)}\ ,\ 1\leq i\leq M-1$
are
free. The other diagonal elements satisfy
\begin{equation}
a^{\alpha}_{ik}=\sum_{p=i}^{k-1} a^{\alpha}_{p(p+1)}\ \ \ \ \ k>i+1
\label{eq:diag}
\end{equation}
where we have introduced the compact notation $A_{ik\,,\,ik}^{\alpha}\equiv
a_{ik}^{\alpha}$.
\end{description}

The canonical forms of the characteristic matrices $A^{\alpha}$ and the
constants
$\sigma^{\alpha\beta}$ satisfy the following conditions:
\begin{description}
\item[1.] The set of matrices $A^{\alpha}$ have the form specified above and
are
linearly nilindependent. For $f\geq 2$ they all commute, i.e.
\begin{equation}
\left[A^{\alpha},A^{\beta}\right]=0.
\label{eq:AA}
\end{equation}
\item[2.] All constants $\sigma^{\alpha\beta}$ vanish unless we have
$a^{\gamma}_{1M}=0\ $ for $\gamma=1,\ldots,f$ simultaneously for all $\gamma$.
\item[3.] The remaining off-diagonal elements $A^{\alpha}_{ik\,,\,ab}$ also
vanish,
unless the diagonal elements satisfy $a^{\beta}_{ik}=a^{\beta}_{ab}\ $ for
$\beta=1,\ldots,f$ simultaneously for all $\beta$.
\item[4.] The maximal number of non-nilpotent elements is $f_{max}=M-1$ and in
this
case the non-nilpotent elements always commute, i.e.

\begin{equation}
\left[X^{\alpha},X^{\beta}\right]=0.
\label{eq:XX=0}
\end{equation}
Furthermore, the characteristic matrices $A^{\alpha}$ are explicitly given by
the diagonal
matrices

\begin{equation}
a^{\alpha}_{ik}=\sum_{p=i}^{k-1}\delta_{\alpha,p}\ \ \ \ \
1\leq i<k\leq M\ \ \ \ 1\leq \alpha\leq M-1.
\label{eq:An-1}
\end{equation}
\item[5.] For $f=1$ the matrix $A$ has at most $M-2$ off-diagonal elements
that
can
be normalized to $+1$ for $K=\mathbb{C}$ and to $+1$, or $-1$ for
$K=\mathbb{R}$.
\end{description}

\subsection {Differential operators and the system of equations}

Using the preceding results, we can construct (as in Section~2) the
differential
operators realizing a basis for the coadjoint representation of the Lie
algebras
$L(M,f)$:

\begin{eqnarray}
\label{eq:N*}
\hat N_{ik}&=&\sum_{b=k+1}^{M} n_{ib}\frac{\partial}{\partial
n_{kb}}-\sum_{a=1}^{i-1} n_{ak}\frac{\partial}{\partial
n_{ai}}-\sum_{\alpha=1}^{f}\left(a^{\alpha}_{ik}n_{ik}+
\Gamma^{\alpha}_{ik}\right)
\frac{\partial}{\partial x^{\alpha}} \\*[2ex]
\label{eq:X}
\hat X^{\alpha}&=&\sum_{i<k}\left(a_{ik}^{\alpha}n_{ik}+
\Gamma^{\alpha}_{ik}\right)\frac{\partial}{\partial
n_{ik}}+
\sum_{\beta=1}^{f}\left(\sigma^{\alpha\beta}n_{1N}\right)\frac{\partial}
{\partial x^{\beta}}.
\end{eqnarray}
We have introduced the notation

\begin{equation}
\begin{array}{rll}
\Gamma_{12}^{\alpha}&\equiv& A^{\alpha}_{12,2M}\,n_{2M} \\
\Gamma_{j(j+1)}^{\alpha} &\equiv&  A^{\alpha}_{j(j+1)\,,\,1M}\,n_{1M}\;\;\; \;\;
j=2,3,\ldots,M-2 \\
\Gamma_{(M-1)M}^{\alpha}&\equiv& A^{\alpha}_{(M-1)M\,,\,1(M-1)}\,n_{1(M-1)}
\\
\Gamma_{lm}^{\alpha} &\equiv&  0\;\;\;\;\; m-l\geq 2.
\end{array}
\end{equation}

In the generic case the differential operators
(\ref{eq:X})
will not contain the second summation since $\sigma^{\alpha\beta}=0$ unless
$a^{\gamma}_{1M}=0$ for $\gamma=1,\ldots,f$.

Equation (\ref{eq:YF}) determining the invariants in our case amounts to the
system
of equations

\begin{eqnarray}
\hat N_{ik}\cdot F(n_{12}, n_{23},\ldots, n_{1M}, x^{1},\ldots,
x^{f})&=&0\;\;\;\;\;\; 1\leq i<k \leq M
\label{eq:NF2} \\*[2ex]
\hat X^{\alpha}\cdot F(n_{12}, n_{23},\ldots, n_{1M}, x^{1},\ldots,
x^{f})&=&0\;\;\;\;\;\; \alpha=1,\ldots, f.
\label{eq:XF}
\end{eqnarray}

 It is useful to construct linear combinations of these operators that involve
only
$x$ derivatives. These linear combinations are not elements of the Lie
algebra
$L(M,f)$, since they have variable coefficients. This is permitted since we
are
now
treating equations (\ref{eq:NF2}) and (\ref{eq:XF}) simply as a system of
linear
partial differential equations.

Let us associate a differential operator $\hat Z_{\mu}$ with each invariant
$Z_{\mu}$
of the nilpotent Lie algebra $T(M)$ (see eq.(\ref{eq:Z})). For each $Z_{\mu}$
we take a
sum of
$\mu$ determinants of the form (\ref{eq:Z}) and in each of them we replace one
column
of scalars by a column of operators $\hat N_{ik}$. For examples, we have

\begin{equation}
\begin{array}{l}
\hat Z_{1}=\hat N_{1M}\;\;\;\;\;\;\hat Z_{2}=\left|
\begin{array}{cc}
\hat N_{1(M-1)} & n_{1M} \\
\hat N_{2(M-1)} & n_{2M}
\end{array}
\right| + \left|
\begin{array}{cc}
n_{1(M-1)} & \hat N_{1M} \\
n_{2(M-1)} & \hat N_{2M}
\end{array}
\right| \\*[2ex]
%\hat Z_{3}= \left|
%\begin{array}{ccc}
%\hat N_{1(M-2)} & n_{1(M-1)} & n_{1M} \\
%\hat N_{2(M-2)} & n_{2(M-1)} & n_{2M} \\
%\hat N_{3(M-2)} & n_{3(M-1)} & n_{3M}
%\end{array}
%\right| + \left|
%\begin{array}{ccc}
%n_{1(M-2)} & \hat N_{1(M-1)} & n_{1M} \\
%n_{2(M-2)} & \hat N_{2(M-1)} & n_{2M} \\
%n_{3(M-2)} & \hat N_{3(M-1)} & n_{3M}
%\end{array}
%\right| + \left|
%\begin{array}{ccc}
%n_{1(M-2)} & n_{1(M-1)} & \hat N_{1M} \\
%n_{2(M-2)} & n_{2(M-1)} & \hat N_{2M} \\
%n_{3(M-2)} & n_{3(M-1)} & \hat N_{3M}
%\end{array}
%\right| \;,
\end{array}
\end{equation}
and in general, we have the formula
\begin{equation}
\hat Z_{\mu}=\sum_{j=1}^{\mu}
\left|
\begin{array}{cccccc}
n_{1(M-\mu+1)} & n_{1(M-\mu+2)} & \cdots & \hat N_{1(M-\mu+j)}& \cdots &
n_{1M}
\\
n_{2(M-\mu+1)} & n_{2(M-\mu+2)} & \cdots & \hat N_{2(M-\mu+j)}& \cdots &
n_{2M}
\\
\vdots & \vdots & & \vdots & & \vdots \\
n_{\mu(M-\mu+1)} & n_{\mu(M-\mu+2)} & \cdots & \hat N_{\mu(M-\mu+j)}& \cdots &
n_{\mu\,M}
\end{array}
\right|\;\;\;\; 1\leq \mu\leq \left[\frac{M}{2}\right]\; .
\label{eq:opZ}
\end{equation}
It is a straightforward calculation to prove that we have

\begin{equation}
\hat Z_{\mu}=\sum_{\alpha=1}^{f}f_{\alpha}(n_{ik})\frac{\partial}{\partial
x^{\alpha}}
\end{equation}
i.e. that all the $n_{ik}$ derivatives drop out. For example, when the
stucture matrices $A^{\alpha}$ are diagonal we obtain the formula

\begin{equation}
\hat
Z_{j}=-Z_{j}\left(\sum_{\alpha=1}^{f}\sum_{\mu=1}^{j}a_{\mu(M-\mu+1)}^{\alpha}
\frac{\partial}{\partial x^{\alpha}}\right).
\label{eq:^Z}
\end{equation}
\emph{Remark}: For non-diagonal matrices $A^{\alpha}$, this formula is generic
 for
odd $M$. However, for even $M$, off-diagonal terms will appear.

We can construct $[M/2]$ such operators; at most $f$ of
them
are
linearly independent.

\subsection{Examples: Invariants of $L(4,f)$}

Let us now illustrate the procedure to obtain the functionally independent
invariants
for the solvable Lie algebras $L(4,f)$, $f=1,2$ or $3$. For each algebra
$L(4,f)$ we
will state results concerning the form and the number of invariants. For each
Lemma,
the
strategy that we will adopt to prove it is the following.

We will separate the proof in two parts:

\begin{description}
\item[(A)] We find the invariants depending only on the variables
$n_{ab}\;,\;1\leq a<b\leq 4$.
\item[(B)] We find the invariants which are dependent on variables $n_{ik}$
\textbf{and} $x^{\alpha}\;,\;\alpha=1,\ldots,f$.
\end{description}

In each of these cases, we will apply the differential operators $\hat
N_{ik}$
and $\hat X^{\alpha}$ of the coadjoint representation of $L(4,f)$, on the
functions
$F=F(\{n_{ab}\},\{x^{\alpha}\})$.
However in the case (A), since we postulate that the functions $F$ only depend
on the
variables
$n_{ab}$, the differential operators $\hat N_{ik}$ will be the same as the
operators
of the nilpotent Lie algebra $T(4)$ (the $x$ derivatives do not act on $F$).
Therefore, by using the results of
Theorem~\ref{theo:I=Z}, we will only have to apply the differential operators
$X^\alpha$ on functions of the type

\begin{equation}
F=F(Z_{1},Z_{2})
\end{equation}
where $Z_{1}=n_{14}$ and $Z_{2}=n_{13}\,n_{24}-n_{23}\,n_{14}$.

In the case (B), we will begin by imposing
\begin{equation}
\hat Z_{j}\cdot F(n_{ab}\,,\,x^{\alpha})=0\;\;\;\;\;\; j=1,2
\end{equation}
such that the dependence on the $x^{\alpha}$ variables is preserved in $F$.
Then
we
will apply all the differential operators (\ref{eq:N*}) and (\ref{eq:X}) of
the
coadjoint representation of $L(4,f)$.

\subsubsection{The Lie algebras $L(4,1)$}

The characteristic matrix $A$ of these Lie algebras $L(4,1)$ has the form
\cite{1}

\begin{equation}
A=\left(
\begin{array}{cccccc}
a_{12}&0&0&0&\lambda_{1}&0 \\
&a_{23}&0&0&0&\lambda_{2} \\
&&a_{34}&\lambda_{3} &0&0 \\
&&& a_{13} &0&0\\
&&&& a_{24} & 0\\
&&&&& a_{14} \\
\end{array}
\right)
\label{eq:AL41}
\end{equation}
where we have at most $2$ non-zero off-diagonal elements $\lambda_{i}$ and by
eq.(\ref{eq:diag}) $a_{13}$, $a_{24}$ and $a_{14}$ are determined in
terms of $a_{12}$, $a_{23}$ and $a_{34}$.

\newtheorem{lemme1}{Lemma}
\begin{lemme1}
A solvable triangular Lie algebra of the type $L(4,1)$ has either $3$
invariants, or
$1$ invariant.
\begin{description}
\item[1)] Three invariants exist iff the conditions

\begin{equation}
a_{14}=a_{23}=\lambda_{2}=0
\label{eq:a14a23}
\end{equation}
are satisfied. In this case the algebra can be characterized by
$a_{12}=-a_{34}=1\;,\;a_{23}=0\;,\;\lambda_{1}=\lambda_{2}=\lambda_{3}=0$ in
characteristic matrix $(\ref{eq:AL41})$. A basis for the invariants is:

\begin{eqnarray}
I_{1}&=&Z_{1}
\label{eq:theo21I1} \\
I_{2}&=&Z_{2}
\label{eq:theo21I2} \\
I_{3}&=&(n_{12}\,n_{24}+n_{13}\,n_{34})+n_{14}\,x.
\label{eq:theo21I3}
\end{eqnarray}
Otherwise there exists precisely one invariant. Two types of Lie algebras
occur.
\item[2)] $(a_{12}+a_{34},a_{23})\neq (0,0)$ and $\lambda_{2}=0$ in matrix
$(\ref{eq:AL41})$. The invariant is:

\begin{equation}
I=\frac{(Z_{2})^{a_{14}}}{(Z_{1})^{a_{14}+a_{23}}}.
\label{eq:theo22}
\end{equation}
\item[3)] $a_{12}+a_{34}=0,\;\lambda_{2}=1,\;\;a_{23}$ is a free parameter in
matrix $(\ref{eq:AL41})$ and the invariant is:
\begin{equation}
I=a_{23}\frac{Z_{2}}{(Z_{1})^{2}}-\ln Z_{1}.
\label{eq:theo23}
\end{equation}
\end{description}
\label{theo:L41}
\end{lemme1}

\emph{Proof.}

\textbf{(A)} We impose that the differential operator $\hat X$ of
eq.(\ref{eq:X}) should annihilates the functions of type
$F=F(Z_{1},Z_{2})$, i.e.

\begin{equation}
\begin{array}{lll}
\hat X\cdot F&=&\left[
(a_{12}\,n_{12}+\lambda_{1}\,n_{24})\displaystyle\frac{\partial}{\partial
n_{12}}
+(a_{23}\,n_{23}+\lambda_{2}\,n_{14})\displaystyle\frac{\partial}{\partial
n_{23}}\right.
\\*[2ex]
&&\left.+(a_{34}\,n_{34}+
\lambda_{3}\,n_{13})\displaystyle\frac{\partial}{\partial n_{34}}
+a_{13}\,n_{13}\displaystyle\frac{\partial}{\partial n_{13}}
+a_{24}\,n_{24}\displaystyle\frac{\partial}{\partial n_{24}} \right. \\
&&\left.+a_{14}\,n_{14}\displaystyle\frac{\partial}{\partial n_{14}}
\right]F
\\*[2ex]
&=& a_{14}\,Z_{1}\,\displaystyle\frac{\partial F}
{\partial
Z_{1}}+\left[(a_{14}+a_{23})\,Z_{2}-\lambda_{2}\,(Z_{1})^{2}\right]\,
\displaystyle\frac{\partial F}{\partial Z_{2}}=0.
\end{array}
\label{eq:XF2}
\end{equation}

We first note that if we have $a_{14}=a_{23}=\lambda_{2}=0$, i.e. conditions
(\ref{eq:a14a23}) which implies $a_{12}+a_{34}=0$ from eq.(\ref{eq:diag}), then
both $Z_{1}$ and
$Z_{2}$
are
invariants. Also, the matrix $A$ can, with no loss of generality \cite{1}, be
diagonalized and set equal to
\begin{equation}
\begin{array}{llllll}
A=\diag(1& 0 & -1 & 1 & -1 & 0).
\end{array}
\label{A1}
\end{equation}
In all other cases eq.(\ref{eq:XF2}) implies that just one invariant of this
type
exists. We obtain it using the method of characteristics.

Two cases arise:
\begin{description}
\item[(i) $\lambda_{2}=0$]: The invariant is then given by (\ref{eq:theo22}),
with
$(a_{12}+a_{34},a_{23})\neq (0,0)$.
\item[(ii) $\lambda_{2}\neq 0$]: From our previous article \cite{1}, we know
that in
this case we can normalize $\lambda_{2}$ to $1$ and we necessarily have
$a_{23}=a_{14}$, which implies $a_{12}+a_{34}=0$. Hence, we obtain the
invariant
(\ref{eq:theo23}), where $a_{23}$ is a free parameter.
\end{description}

\textbf{(B)} In this case we impose $\hat Z_{j}\cdot F=0$ ($j=1,2$) for
functions of the type
$F=F(n_{12},n_{23},n_{34},n_{13},n_{24},n_{14},x)$ and the differential
 operators $\hat{Z}_{j}$ are given by

\begin{eqnarray}
\hat Z_{1}&\equiv& \hat N_{14}=-a_{14}\,Z_{1}\frac{\partial}{\partial x} \\
\hat Z_{2}&\equiv&n_{13}\,\hat N_{24}-n_{23}\,\hat N_{14}
+n_{24}\,\hat N_{13}-n_{14}\,\hat N_{23} \nonumber \\*[2ex]
&=&\left[-(a_{14}+a_{23})\, Z_{2}+\lambda_{2}\,(Z_{1})^{2}\right]
\frac{\partial}{\partial x}.
\end{eqnarray}

Hence the required dependence on $x$ will survive only if we have
$a_{14}=a_{23}=\lambda_{2}=0$. This coincides with eq.(\ref{eq:a14a23}), the
condition for $Z_{1}$ and $Z_{2}$ to be invariant. Furthermore, we can
normalize
$a_{12}$ to $1$ and cancel $\lambda_{1}$ and $\lambda_{3}$ by transformations
\cite{1}.

We now apply all the differential operators of the coadjoint representation of
$L(4,1)$ and the final result is that we obtain two invariants
(\ref{eq:theo21I1})
and (\ref{eq:theo21I2}) independent of $x$ and one invariant
(\ref{eq:theo21I3})
depending on $x$. $\Box$

\subsubsection{The Lie algebras $L(4,2)$}
The Lie algebras $L(4,2)$ have the following characteristic matrices \cite{1}:
\begin{equation}
A^{1}=\left(
\begin{array}{cccccc}
a_{12}&&&&& \\
&a_{23}&&&& \\
&&a_{34}&&& \\
&&&a_{13}&& \\
&&&&a_{24}& \\
&&&&&a_{14} \\
\end{array}
\right)\;\;
A^{2}=\left(
\begin{array}{cccccc}
b_{12}&0&0&0&\lambda_{1}&0 \\
&b_{23}&0&0&0&\lambda_{2} \\
&&b_{34}&\lambda_{3}&0&0 \\
&&&b_{13}&0&0 \\
&&&&b_{24}&0 \\
&&&&&b_{14} \\
\end{array}
\right)
\label{eq:AL42}
\end{equation}
where we have at most one off-diagonal element in $A^{2}$ and $a_{ik},\,
b_{ik}$
satisfy the eq.(\ref{eq:diag}). Furthermore, the coefficient $\sigma^{12}$ in
eq.(\ref{eq:xx}) is in the generic case zero (i.e. the two non-nilpotent
elements
commute). However, for the particular case $a_{14}=0=b_{14}$, we can have
$\sigma^{12}\neq 0$ in eq.(\ref{eq:xx}).
\newtheorem{lemme2}[lemme1]{Lemma}
\begin{lemme2}
A solvable triangular Lie algebra of the type $L(4,2)$ has either $2$
invariants or none. Two invariants exist iff the conditions

\begin{eqnarray}
b_{23}\,(a_{12}+a_{34})-a_{23}\,(b_{12}+b_{34})&=&0
\label{eq:cond1} \\
a_{14}\,\lambda_{2}&=&0\;
\label{eq:cond2}
\end{eqnarray}
are satisfied simultaneously. They lead to the following algebras and
invariants.
\begin{description}

\item[1)] $a_{12}=-a_{34}=b_{23}=\lambda_{2}=1$ and
$a_{23}=b_{12}=b_{34}=\lambda_{1}=\lambda_{3}=\sigma^{12}=0$ in
matrices $(\ref{eq:AL42})$ and
a basis for the invariants is:

\begin{eqnarray}
I_{1}&=&\frac{Z_{2}}{(Z_{1})^{2}}+\ln Z_{1}
\label{eq:theo31I1}\\*[2ex]
I_{2}&=&\frac{n_{12}\,n_{24}+n_{13}\,n_{34}}{n_{14}}+x^{1}.
\label{eq:theo31I2}
\end{eqnarray}

\item[2a)] $a_{12}=-a_{34}=b_{23}=1\;,\;
a_{23}=b_{12}=\lambda_{1}=\lambda_{2}=\lambda_{3}=\sigma^{12}=0$ and $b_{34}$
a
free
parameter in matrices $(\ref{eq:AL42})$,
\item[2b)] $a_{12}=b_{34}=1$ and
$a_{23}=a_{34}=b_{12}=b_{23}=\lambda_{1}=\lambda_{2}=\lambda_{3}=
\sigma^{12}=0$
in matrices $(\ref{eq:AL42})$

In both cases we have the invariants:

\begin{eqnarray}
I_{1}&=&\frac{(Z_{2})^{a_{14}}}{(Z_{1})^{a_{14}+a_{23}}}
\label{eq:theo32I1} \\*[2ex]
I_{2}&=&
(a_{34}b_{13}-b_{34}a_{13})\left(\frac{n_{12}n_{24}+
n_{13}n_{34}}{n_{14}}\right)+a_{14}x^{2}-b_{14}x^{1}.
\label{eq:theo32I2}
\end{eqnarray}

\item[3)] $a_{12}=-a_{34}=b_{23}=-b_{34}=1$ and
$a_{23}=b_{12}=\lambda_{1}=\lambda_{2}=\lambda_{3}=0$ in matrices
$(\ref{eq:AL42})$ and
the
invariants are:

\begin{eqnarray}
I_{1} &=& Z_{1}
\label{eq:theo33I1}\\*[2ex]
I_{2} &=&
n_{12}\,n_{24}+n_{13}\,n_{34}+Z_{1}\,x^{1}+\sigma^{12}\,(Z_{1})^{2}\,\ln
Z_{2}\;.
\label{eq:theo33I2}
\end{eqnarray}

Otherwise, there is no invariant.
 \end{description}
\label{theo:L42}
\end{lemme2}

\emph{Proof.}

\textbf{(A)} We first apply differential operators $\hat X_{1}$ and
$\hat X_{2}$
on
functions of type $F=F(Z_{1},Z_{2})$. We obtain a system of two linear partial
differential
equations given by

\begin{equation}
\left(
\begin{array}{c}
\hat X^{1}\cdot F \\*[2ex]
\hat X^{2}\cdot F
\end{array}
\right)=\left(
\begin{array}{cc}
a_{14}\,Z_{1} & (a_{14}+a_{23})\,Z_{2} \\*[2ex]
b_{14}\,Z_{1} & (b_{14}+b_{23})\,Z_{2}-\lambda_{2}\,(Z_{1})^{2}
\end{array}
\right)\left(
\begin{array}{c}
\displaystyle\frac{\partial F}{\partial Z_{1}} \\*[2ex]
\displaystyle\frac{\partial F}{\partial Z_{2}}
\end{array}
\right)=0.
\label{eq:X1FX2F}
\end{equation}

The rank of the $2\times 2$ matrix in eq.(\ref{eq:X1FX2F}) cannot be zero,
since
then
matrices $A^{1}$ and $A^{2}$ would not be linearly nilindependent. Also, if
the
rank
is $2$ there is no invariant that depends only on $Z_{1}$ and $Z_{2}$.
However,
solution exist if the rank of the matrix is $1$ for all values of $Z_{1}$ and
$Z_{2}$. This gives conditions (\ref{eq:cond1}) and (\ref{eq:cond2}).

Let us now assume that the condition (\ref{eq:cond1}) is respected. We
consider
the
diagonal and the non-diagonal cases separately.

\begin{description}
\item[(i) $\lambda_{2}=0$]: In this case, we obtain the invariant
(\ref{eq:theo22})
for $(a_{12}+a_{34},a_{23})\neq (0,0)$.

\item[(ii) $\lambda_{2}\neq 0\;,\; a_{14}=0$]: Since $\lambda_{2}$ is non-zero
in
$A^{2}$, we necessarily have $b_{23}=b_{14}$, i.e. $b_{12}+b_{34}=0$ which
gives
the
condition $a_{23}\,b_{23}=0$ by (\ref{eq:cond1}). Two cases are possible under
these
condition.

One case gives the invariant (\ref{eq:theo31I1}) for
$a_{12}=-a_{34}=b_{23}=\lambda_{2}=1$ and
$a_{23}=b_{12}=b_{34}=\lambda_{1}=\lambda_{3}=0$.

In the other case, we simply obtain the invariant $I=Z_{1}$ for the Lie
algebra
characterized by $a_{23}=b_{12}=-b_{34}=\lambda_{2}=1$,
$b_{23}=\lambda_{1}=\lambda_{3}=0$ and $a_{34}=-(a_{12}+1)$ (with $a_{12}$ a
free
parameter).

\emph{Remark.} The case $a_{23}=0=b_{23}$ gives two {\em nildependent}
matrices
$A^{1},\;A^{2}$ and is therefore not considered.
\end{description}

\textbf{(B)} In this case, we begin by applying the differential operators
 $\hat Z_{1},\;\hat Z_{2}$ on functions of type
$F=F(n_{12},n_{23},n_{34},n_{13},n_{24},n_{14},x^{1},x^{2})$, i.e.

\begin{equation}
\left(
\begin{array}{c}
\hat Z_{1}\cdot F\\*[2ex]
\hat Z_{2}\cdot F
\end{array}
\right)=\left(
\begin{array}{cc}
-a_{14}\,Z_{1} & -b_{14}\,Z_{1} \\*[2ex]
(a_{23}+a_{14})\,Z_{2} & (b_{23}+b_{14})\,Z_{2}-\lambda_{2}\,(Z_{1})^{2}
\end{array}
\right)\left(
\begin{array}{c}
\displaystyle\frac{\partial F}{\partial x^{1}} \\*[2ex]
\displaystyle\frac{\partial F}{\partial x^{2}}
\end{array}
\right)=0\;.
\label{eq:Z1Z2}
\end{equation}
The dependence on $x^{1}$ and $x^{2}$ can exist only if the determinant of the
$2\times 2$ matrix in (\ref{eq:Z1Z2}) is zero. This again imposes the
conditions
(\ref{eq:cond1}) and (\ref{eq:cond2}).

Let us again assume that the condition (\ref{eq:cond1}) is satisfied. We
separate the
problem into three distinct cases.
\begin{description}
\item[(i) $(a_{14},b_{14})\neq (0,0)\;,\;\lambda_{2}\neq 0$]: The condition
$\lambda_{2}\neq 0$ implies two consequences. First we have from
(\ref{eq:cond2})
that $a_{14}=0$ and therefore $b_{14}\neq 0$. Second, we necessarily have
$b_{23}=b_{14}$ which implies from (\ref{eq:cond2}) that $b_{23}\,a_{23}=0$

In this case, the invariants are (\ref{eq:theo31I1}) and (\ref{eq:theo31I2})
and
the
Lie algebra $L(4,2)$ satisfies
$a_{12}=-a_{34}=b_{23}=\lambda_{2}=1\;,
\;a_{23}=b_{12}=b_{34}=\lambda_{1}=\lambda_{3}=\sigma^{12}=0$.

\item[(ii) $(a_{14},b_{14})\neq (0,0)\;,\;\lambda_{2}=0$]: In this case, two
triangular solvable Lie algebras are associated with the invariants
(\ref{eq:theo32I1}) and (\ref{eq:theo32I2}). One Lie algebra is characterized
by
the
parameters $a_{12}=-a_{34}=b_{23}=1\;,\;
a_{23}=b_{12}=\lambda_{1}=\lambda_{2}=\lambda_{3}=\sigma^{12}=0$ and $b_{34}$
a
free
parameter. The other Lie algebra is characterized by $a_{12}=b_{34}=1$ and
$a_{23}=a_{34}=b_{12}=b_{23}=\lambda_{1}=\lambda_{2}=\lambda_{3}=
\sigma^{12}=0$.

\item[(iii) $(a_{14},b_{14})=(0,0)$]: In this case, we see that conditions
(\ref{eq:cond1}) and (\ref{eq:cond2}) are automatically respected. Also, we
can
have
a non-zero $\sigma^{12}$ in eq.(\ref{eq:X}).

Since $a_{14}=0=b_{14}$, we can substitute $a_{34}$ by $-(a_{12}+a_{23})$ and
$b_{34}$ by $-(b_{12}+b_{23})$ in the characteristic matrices (\ref{eq:AL42}).
However, by
imposing the commutativity (\ref{eq:AA}) and the nilindependence of the
matrices
$A^{1}$ and $A^{2}$, we obtain $a_{12}=-a_{34}=b_{23}=-b_{34}=1$ and
$a_{23}=b_{12}=\lambda_{1}=\lambda_{2}=\lambda_{3}=0$. Hence, we obtain the
two
invariants (\ref{eq:theo33I1}) and (\ref{eq:theo33I2}). $\Box$

\end{description}

\subsubsection{The Lie algebra $L(4,3)$}
For the Lie algebra $L(4,3)$, we have diagonal characteristic matrices
given
by

\begin{equation}
\begin{array}{lll}
A^{1}&=&\diag
\left(
\begin{array}{cccccc}
1&0&0&1&0&1
\end{array}
\right)\;
A^{2}=\diag
\left(
\begin{array}{cccccc}
0&1&0&1&1&1
\end{array}
\right) \\*[2ex]
A^{3}&=&\diag
\left(
\begin{array}{cccccc}
0&0&1&0&1&1
\end{array}
\right).
\end{array}
\label{eq:AL43}
\end{equation}
Furthermore, the non-nilpotent elements commute, i.e.
$\sigma^{\alpha\,\beta}=0,\;\alpha,\,\beta=1,2,3$ (see equations
(\ref{eq:xx}) and
(\ref{eq:XX=0}))

\newtheorem{lemme3}[lemme1]{Lemma}
\begin{lemme3}
The triangular solvable Lie algebra $L(4,3)$  has precisely $1$ invariant given
by
\begin{equation}
I=\frac{n_{12}\,n_{24}+n_{13}\,n_{34}}{n_{14}}+(x^{1}-x^{3})\;.
\label{eq:theo3I}
\end{equation}
\label{theo:L43}
\end{lemme3}

\emph{Proof.}

\textbf{(A)} In this case, it is easy to demonstrate that after we have
applied
the
differential operator $\hat X^{1}$ on functions of type $F=F(Z_{1},Z_{2})$,
we obtain
the
quotient of $Z_{2}$ over $Z_{1}$. However, when we apply operator $\hat X^{2}$
on
functions $\tilde F=\tilde F(I)$ with $I=Z_{2}/Z_{1}$, we obtain

\begin{equation}
\begin{array}{lll}
0&=&\hat X^{2}\cdot \tilde F=\left(
n_{23}\,\displaystyle\frac{\partial}{\partial
n_{23}}+n_{13}\,\displaystyle\frac{\partial}{\partial n_{13}}+
n_{24}\,\displaystyle\frac{\partial}{\partial
n_{24}}+n_{14}\,\displaystyle\frac{\partial}{\partial n_{14}}\right)\tilde F
\\*[2ex]
&=& I\displaystyle\frac{\partial}{\partial I}\tilde F\;.
\end{array}
\end{equation}
Therefore, there is no invariant in this case.

\textbf{(B)} We first impose that the differential operators $\hat Z_{1}$ and
$\hat
Z_{2}$ annihilate the functions of type
$F=F(n_{12},n_{23},n_{34},n_{13},n_{24},n_{14},x^{1},x^{2},x^{3})$, where

\begin{eqnarray}
\hat Z_{1}&=&-Z_{1}\left(\frac{\partial}{\partial
x^{1}}+\frac{\partial}{\partial
x^{2}}+\frac{\partial}{\partial x^{3}}\right) \\
\hat Z_{2} &=& -Z_{2}\left(\frac{\partial}{\partial
x^{1}}+2\frac{\partial}{\partial
x^{2}}+\frac{\partial}{\partial x^{3}}\right).
\end{eqnarray}
Since the Lie algebra $L(4,3)$ has no parameters, these conditions are not on
the
parameters of the algebra (as before) but on the $x$ dependence of the
invariant. Hence, the new functions on which we will
apply
all
the differential operators of the coadjoint representation of $L(4,3)$ are of
the type
$\tilde F=\tilde F(n_{12},n_{23},n_{34},n_{13},n_{24},n_{14},x^{1}-x^{3})$.
We then
obtain the invariant (\ref{eq:theo3I}) by imposing that the operators of the
coadjoint representation of $L(4,3)$ annihilate $\tilde F$. $\Box$

\subsection{General results}

\newtheorem{prop1}{Proposition}
\begin{prop1}
The triangular solvable Lie algebra $L(M,M-1)$ has precisely
$\left[\frac{M-1}{2}\right]$ functionally independent invariants. A basis is
given by
\begin{equation}
I_{\mu}=\frac{(-1)^{\mu +
1}}{Z_{\mu}}\left(\sum_{\rho=1}^{M-2\mu}W_{\rho}^{(\mu)}
\right)+(x^{\mu}-x^{M-\mu})
\label{eq:I=W+x}
\end{equation}
for $\mu=1,\ldots,\left[\frac{M-1}{2}\right]$. The function $Z_{\mu}$ is the
determinant given by eq.$(\ref{eq:Z})$ and $W_{\rho}^{(\mu)}$ is also a
determinant
function given by the determinant of the $(\mu+1)\times (\mu+1)$ matrix:
\begin{equation}
W_{\rho}^{(\mu)}=\left|
\begin{array}{ccccc}
n_{1(\rho+\mu)} & n_{1(M-\mu+1)} & n_{1(M-\mu+2)} & \cdots & n_{1M}  \\
n_{2(\rho+\mu)} & n_{2(M-\mu+1)} & n_{2(M-\mu+2)} & \cdots & n_{2M}  \\
\vdots & \vdots & \vdots & & \vdots \\
n_{\mu(\rho+\mu)} & n_{\mu(M-\mu+1)} & n_{\mu(M-\mu +2)} & \cdots & n_{\mu M}
\\
0 & n_{(\rho+\mu)(M-\mu+1)} & n_{(\rho+\mu)(M-\mu+2)} & \cdots &
n_{(\rho+\mu)M}
\end{array}
\right|.
\label{eq:W}
\end{equation}
\label{theo:L(M,M-1)}
\end{prop1}

\newtheorem{prop2}[prop1]{Proposition}
\begin{prop2}
A diagonal solvable Lie algebra of the type $L(M,1)$ has
$\left[\frac{M}{2}\right]\pm 1$ functionally independent invariants.

\begin{description}
\item[1)] $\left[\frac{M}{2}\right]+1$ invariants exist iff the conditions

\begin{equation}
a_{i(i+1)}+a_{(M-i)(M-i+1)}=0\;\;\;\;\; i=1,\ldots,\left[\frac{M}{2}\right]
\label{eq:a+a=0}
\end{equation}
are satisfied. A basis is given by $[M/2]$ invariants
independent
of
$x$
and one invariant depending on $x$:

\begin{eqnarray}
I_{\mu}&=&Z_{\mu}\;\;\;\;\;\;\; \mu=1,\ldots,\left[\frac{M}{2}\right] \\
I_{\left[\frac{M}{2}\right]+1}&=&
\sum_{\mu=1}^{\left[(M-1)/2\right]}\sum_{\rho=1}^{M-2\mu}
\frac{(-1)^{\mu + 1}}{Z_{\mu}}\,a_{\mu(\mu +1)}\,W_{\rho}^{(\mu)}+x
\label{eq:IM/2+1}
\end{eqnarray}
where the function $Z_{\mu}$ and $W_{\rho}^{(\mu)}$ are determinant functions
given by the equations $(\ref{eq:Z})$ and $(\ref{eq:W})$, respectively.

\item[2)] Otherwise there exist precisely $\left[\frac{M}{2}\right]-1$
invariants, all independent of $x$. A basis is given by
\begin{equation}
I_{\mu}=\frac{\left(Z_{\mu+1}\right)^{\alpha}}{\left(Z_{1}\right)^{\beta}}\ \
\ \
 \mu=1,\ldots,\left[\frac{M}{2}\right]-1
\label{eq:prop22}
\end{equation}
with
\begin{equation}
\frac{\alpha}{\beta}=\frac{a_{1M}}{\displaystyle\sum_{k=1}^{\mu+1}
a_{k(M+1-k)}}
\end{equation}
\end{description}
\label{theo:L(M,1)}
where the function $Z_{\mu}$ is the determinant function given by
eq.$(\ref{eq:Z})$.
\end{prop2}
By {\em diagonal} solvable Lie algebra of the type $L(M,1)$ in Proposition~2, we
mean that the characteristic matrix $A$ of (\ref{eq:xn}) is diagonal.

Propositions 1 and 2 each contains two types of information on the invariants:
They give the form of the invariant functions and the number of functionally
independent invariants. It is an easy calculation to prove that the
functions $I_{\mu}$ of Proposition~1 and Proposition~2 are annihilated by the
coadjoint representation (\ref{eq:N*}),  (\ref{eq:X}) of the Lie algebras
$L(M,M-1)$ and $L(M,1)$, respectively. However, it is much more difficult to
establish the number of functionally independent invariants for Proposition~1
and
Proposition~2. The difficulty is to prove that no further invariants exists. One
way of doing that is to calculate the rank of the antisymmetric matrix
$S=S(L(M,M-1))$ and $S=S(L(M,1))$ of the commutation relations for the
corresponding Lie algebra. The number of invariants is then given by the
difference between the dimension of the solvable Lie algebra and the rank of the
matrix $S$ (see eq.(\ref{eq:ni})).

For the Lie algebra $L(M,M-1)$ of dimension $\frac{1}{2}(M-1)(M+2)$, $S$ is
the antisymmetric matrix given by the elements

\begin{equation}
\begin{array}{c}
S=\{[N_{ik},N_{ab}]\ \ [N_{ik},X^{\alpha}]\} \\*[2ex]
1\leq i<k\leq M\ \ \ \ 1\leq a<b\leq M\ \ \  \ \alpha=1,\ldots, M-1
\end{array}
\end{equation}
and for the Lie algebra $L(M,1)$ of dimension $\frac{1}{2}(M^2-M+2)$, the
matrix $S$ is given by the elements

\begin{equation}
\begin{array}{c}
S=\{[N_{ik},N_{ab}]\ \ [N_{ik},X]\} \\*[2ex]
1\leq i<k\leq M\ \ \ \ 1\leq a<b\leq M.
\end{array}
\end{equation}

For example, the antisymmetric matrix $S$ of the $7$-dimensional Lie algebra
$L(4,1)$ is given by

\begin{equation}
S=\left(
\begin{array}{ccccccc}
0 & N_{13} & 0 & 0 & N_{14} & 0 & -a_{12}N_{12}  \\
-N_{13} & 0 & N_{24} & 0 & 0 & 0 & -a_{23}N_{23}  \\
0 & -N_{24} & 0 & -N_{14} & 0 & 0 & -a_{34}N_{34} \\
0 & 0 & N_{14} & 0 & 0 & 0 & -a_{13}N_{13} \\
-N_{14} & 0 & 0 & 0 & 0 & 0 & -a_{24}N_{24} \\
0 & 0 & 0 & 0 & 0 & 0 & -a_{14}N_{14} \\
a_{12}N_{12} & a_{23}N_{23} & a_{34}N_{34} & a_{13}N_{13} & a_{24}N_{24} &
a_{14}N_{14} & 0 \\
\end{array}
\right)
\end{equation}
where the parameters $a_{13}$, $a_{24}$ and $a_{14}$ are given in terms of
$a_{12}$, $a_{23}$ and $a_{34}$ by the relation
(\ref{eq:diag}). Hence, it is easy to calculate that
\begin{equation}
rank(S)=\left\{
\begin{array}{lll}
4 && for\ a_{14}=a_{23}=0 \\*[2ex]
6 && otherwise
\end{array}
\right.
\end{equation}
giving, respectively, three and one invariants (in accordance with Proposition~2
and Lemma~1).

We have calculated the ranks of the matrices $S(L(M,M-1))$ and $S(L(M,1))$ for
$M\leq 13$ and $M\leq 8$, respectively, using the symbolic package MAPLE. We
conjecture that Proposition~1 and 2 hold for all $M$.

\section{Conclusions}

The problem of finding all invariants of the coadjoint representation of the
triangular nilpotent algebras $T(M)$ is solved completely by Theorem~1. A basis
for the invariants consists of polynomials and provides Casimir operators in the
enveloping algebra of $T(M)$.

The situation with the solvable triangular Lie algebras $L(M,f)$ is more
complicated. We have provided guidelines for calculating the invariants for all
values of $M$, but presented comprehensive results only for $M=4$. We have also
presented conjectures concerning the invariants of $L(M,M-1)$ and $L(M,1)$ for
all values of $M$ (and verified them for a large range of values of $M$).

The results for $M=4$ show that all invariants are polynomial only in special
cases. In general, rational, irrational and logarithmic type invariants must be
allowed in any basis of invariants.

\section*{Acknowledgments}
The research of P.W. was supported in part by research grants from NSERC of
Canada and FCAR du Qu\'ebec.


\newpage
\begin{thebibliography}{99}

\bibitem{1} Tremblay S and Winternitz P 1998 {\em J. Phys. A: Math. Gen.} {\bf
31} 789
\bibitem{2} Wigner E P 1939 {\em Ann. Math.} {\bf 40} 149
\bibitem{3} Gell-Mann M 1962 {\em Phys. Rev.} {\bf 125} 1067
\bibitem{4} Okubo S 1962 {\em Prog. Theor. Phys.} {\bf 16} 686
\bibitem{5} Bargmann V 1936 {\em Z. Phys.} {\bf 99} 576
\bibitem{6} Engelfield M J 1972 {\em Group Theory and the Coulomb Problem}
(Wiley, New-York)

\bibitem{7} Abellanas L and Martinez Alonso L 1975 {\em J. Math. Phys.}
{\bf 16} 1580

\bibitem{8} Patera J, Sharp R T, Winternitz P and Zassenhaus H 1976
{\em J. Math. Phys.} {\bf 17} 986
\bibitem{9} Kirillov A 1962 {\em Russian Math. Surveys} {\bf 17} 57
\bibitem{10} Kirillov A 1974 {\em \'El\'ements de la Th\'eorie des
Repr\'esentations} (Moscow: Mir)
\bibitem{11} Casimir H 1931 {\em Proc. R. Acad. Amsterdam} {\bf 34} 844
\bibitem{12} Racah G 1950 {\em Rend. Lincei} {\bf 8} 108; 1951 {\bf 37} 28
\bibitem{13} Racah G 1965 {\em Group Theory and Spectroscopy (Springer Tracts
in Modern Physics 37)} (Berlin: Springer)
\bibitem{14} Gel'fand I M 1950 {\em Mat. Sbornik} {\bf 26} 103
\bibitem{15} Berezin F A 1956 {\em Dokl. Akad. Nauk. SSSR (NS)} {\bf 107} 9;
1957
{\em Trudy Mosk. Mat. Obshch.} {\bf 6} 371; 1963 Trudy Mosk. Mat. Obshch. {\bf
12} 453
\bibitem{16} Perelomov A M and Popov V S 1967 {\em Sov. Math. Dokl.} {\bf 8}
631
\bibitem{17} Gruber B and O'Raifeartaigh L 1964 {\em J. Math. Phys.} {\bf 5}
1976

\bibitem{18} Ndogmo J C 2000 {\em J. Phys. A: Math. Gen.} {\bf 33} 2273
\bibitem{19} Ndogmo J C 1996 {\em Ind. J. Math.} {\bf 38} 149
\bibitem{20} Patera J, Sharp R T, Winternitz P and Zassenhaus H 1976
{\em J. Math. Phys.} {\bf 17} 977

\bibitem{21} Rubin J and Winternitz P 1993 {\em J. Phys. A: Math. Gen.}
{\bf 26} 1123
\bibitem{22} Ndogmo J C and Winternitz P 1994 {\em J. Phys. A: Math. Gen.}
{\bf 27}
405; 1994 {\em J. Phys. A: Math. Gen.} {\bf 27} 2787
\bibitem{23} Hietarinta J 1987 {\em Phys. Rep.} {\bf 147} 87
\bibitem{24} Ramani A, Grammaticos B and Bountis T 1989 {\em Phys. Rep.}
{\bf 180} 159
\end{thebibliography}
\end{document}